\documentclass[fleqn,12pt,twoside]{article}
\usepackage{espcrc1,psfig}
\usepackage{graphicx}
\usepackage[figuresright]{rotating}

\newcommand{\gton}{\mathrel{\lower.5ex \hbox{$\stackrel{> }
 {\scriptstyle \sim}$}}}
\newcommand{\lton}{\mathrel{\lower.5ex \hbox{$\stackrel{< }
 {\scriptstyle \sim}$}}}

\newcommand{\AmS}{{\protect\the\textfont2
  A\kern-.1667em\lower.5ex\hbox{M}\kern-.125emS}}

\title{
Prompt Photon and Inclusive $\pi^0$ Production 
 at RHIC and LHC
}
\author{S. Jeon\address{
        Department of Physics, McGill  University, Montreal,
        QC H3A-2T8, Canada}$^,$\address{
        RIKEN-BNL Research Center, Upton, NY 11973-5000, USA}, 
        J. Jalilian-Marian\address{
        Physics Department, Brookhaven National 
        Laboratory, Upton, NY 11973-5000, USA}, 
        I. Sarcevic\address{
        Department of Physics, University of Arizona, Tucson, Arizona
        85721, USA}\thanks{talk presented by I. Sarcevic.}
}

\begin{document}

\maketitle

\begin{abstract}
We present results for prompt photon and 
inclusive $\pi^0$ production in
p-p and A-A collisions  at RHIC and LHC energies.  We include the full
next-to-leading order radiative corrections and nuclear
effects, such as 
nuclear shadowing 
and parton energy loss.  
 We find the
next-to-leading order corrections to be
large and $p_T$ dependent.  
 We show how measurements of $\pi^0$ production at RHIC and LHC, 
at large $p_T$, can provide valuable information about the nature of 
parton energy loss.  
 We calculate the ratio of prompt photons to neutral pions
 and show that at RHIC energies this ratio
increases with $p_T$ approaching one at $p_T \sim 10$ GeV, due to the large
suppression of $\pi^0$ production. We show that
at the LHC, this ratio has steep $p_T$ dependence and approaches
$10\%$ effect at $p_T \sim 20$ GeV.
\end{abstract}

\section{INTRODUCTION}
In high-energy heavy-ion collisions hard scatterings of partons occur
in the early stages of the reaction, well before a quark-gluon plasma
might have been formed, producing 
 fast partons 
 that propagate through the hot and dense medium  
 and lose their energy.  It has been predicted that parton 
energy loss would result in the suppression 
of pion production in heavy-ion collisions relative to
hadron-hadron collisions \cite{wang}.  
Recent data on inclusive $\pi^0$
production at RHIC energy of $\sqrt s=200$ GeV \cite{phenix} and 
at large $p_T$, $3$GeV$\le p_T \le 8$GeV, 
confirms this prediction, however the observed suppression was found to 
become stronger 
with increasing $p_T$, providing a new challenge for theoretical models.  

In addition to being of special interest for studying parton energy loss 
effects, 
 large-$p_T$ $\pi^0$ mesons
form a significant background for the prompt photons.  
 Theoretical predictions
for the prompt photon production \cite{jos} and for the 
ratio of prompt protons to pions at RHIC and LHC energies are
crucial for studying possible quark-gluon plasma formation via 
photons.  

\section{INCLUSIVE $\pi^0$ AND PROMPT PHOTON PRODUCTION AT RHIC AND LHC}
In perturbative QCD, the inclusive cross section
for pion production in a hadronic collision is given by:

\begin{eqnarray}
E_\pi \frac{d^3\sigma}{d^3p_\pi}(\sqrt s,p_\pi)
&=&
\int dx_{a}\int dx_{b} \int dz \sum_{i,j}F_{i}(x_{a},Q^{2}) 
F_{j}(x_{b},Q^{2}) D_{c/\pi}(z,Q^2_f) E_c
{d^3\hat{\sigma}_{ij\rightarrow c X}\over d^3p_c}
\label{eq:factcs}
\end{eqnarray}

\noindent
where $F_{i}(x,Q^{2})$ is the i-th parton distribution in a nucleon,
$D_{c/\pi}(z,Q^2_f)$ is the pion fragmentation function 
and 
${d^3\hat{\sigma}_{ij\rightarrow c X}/ d^3p_c}$ are parton-parton cross
sections.  
Prompt photon production is obtained using similar expression, except 
 that 
photon production has contributions from 
 direct processes and bremsstrahlung processes, 
where only bremsstrahlung processes are convoluted with photon fragmentation 
function, 
$D_{c/\gamma}(z,Q^2_f)$.  

We calculate inclusive pion production in proton-proton collisions using 
MRS99 parton distributions \cite{mrs}, BKK pion fragmentation function \cite{bkk} 
 and 
we include leading-order, 
$O(\alpha_s^2)$, 
and the next-to-leading order, $O(\alpha_s^3)$, subprocesses \cite{se}.  
Our prediction 
for inclusive $\pi^0$ production at RHIC energy of $\sqrt s=200$ GeV, 
and for $p_T>3$GeV \cite{jjs}, 
 was found to be in excellent 
agreement with PHENIX data \cite{phenix}, indicating that 
perturbative QCD approach is justified 
 in this kinematic region.  

To calculate the inclusive cross section for pion production in heavy
ion collisions, we use Eq. (1) with the parton distributions modified 
to include nuclear shadowing effect \cite{eks98}  
and modified fragmentation function to incorrporate parton energy loss 
\cite{hsw}.  
We consider constant parton energy loss \cite{jjs2} as well as energy-dependent 
 energy loss \cite{jjs}.  
In Fig. 1 we show our results for the ratio of inclusive cross section for 
pion production in Au-Au collisions to 
the one in proton-proton collisions, 
$R_{AA}(p_T)$.  
We find that 
for constant energy loss 
the ratio
increases with $p_T$,
while for the energy-dependent case, $\epsilon=\kappa E$, 
the ratio decreases with $p_T$.  
For $\kappa=0.06$ we find excellent agreement with
the recent PHENIX data \cite{rob}.
In Fig. 1 we also 
show our prediction for $R_{AA}(p_T)$ in case of prompt photon 
production.  
The suppression of prompt photons produced in heavy-ion collisions
at RHIC is much less than $\pi^0$ suppresion.  
This is due to the fact that only
bremsstrahlung processes are affected by the parton energy loss,
which contribute $24\%$ to the cross section at $p_T=3$ GeV and
 $6\%$ at $p_T=12$ GeV.
For the same reason, prompt photons are not very sensitive to a different
choice of parton energy loss.  

In Fig. 2 we show the ratio of prompt photon and $\pi^0$ cross sections 
 at $\sqrt s=200$ GeV.  We find 
that for constant energy loss this ratio increases slowly with $p_T$, similar to 
the p-p case, while for $\epsilon=0.06E$, the ratio has strong $p_T$ dependence, 
approaching one at $p_T \sim 10$GeV.  This is due to the strong suppression of 
$\pi^0$ production at large $p_T$.
\begin{figure}[htb]
\begin{minipage}[t]{75mm}
\includegraphics[angle=0,width=75mm]{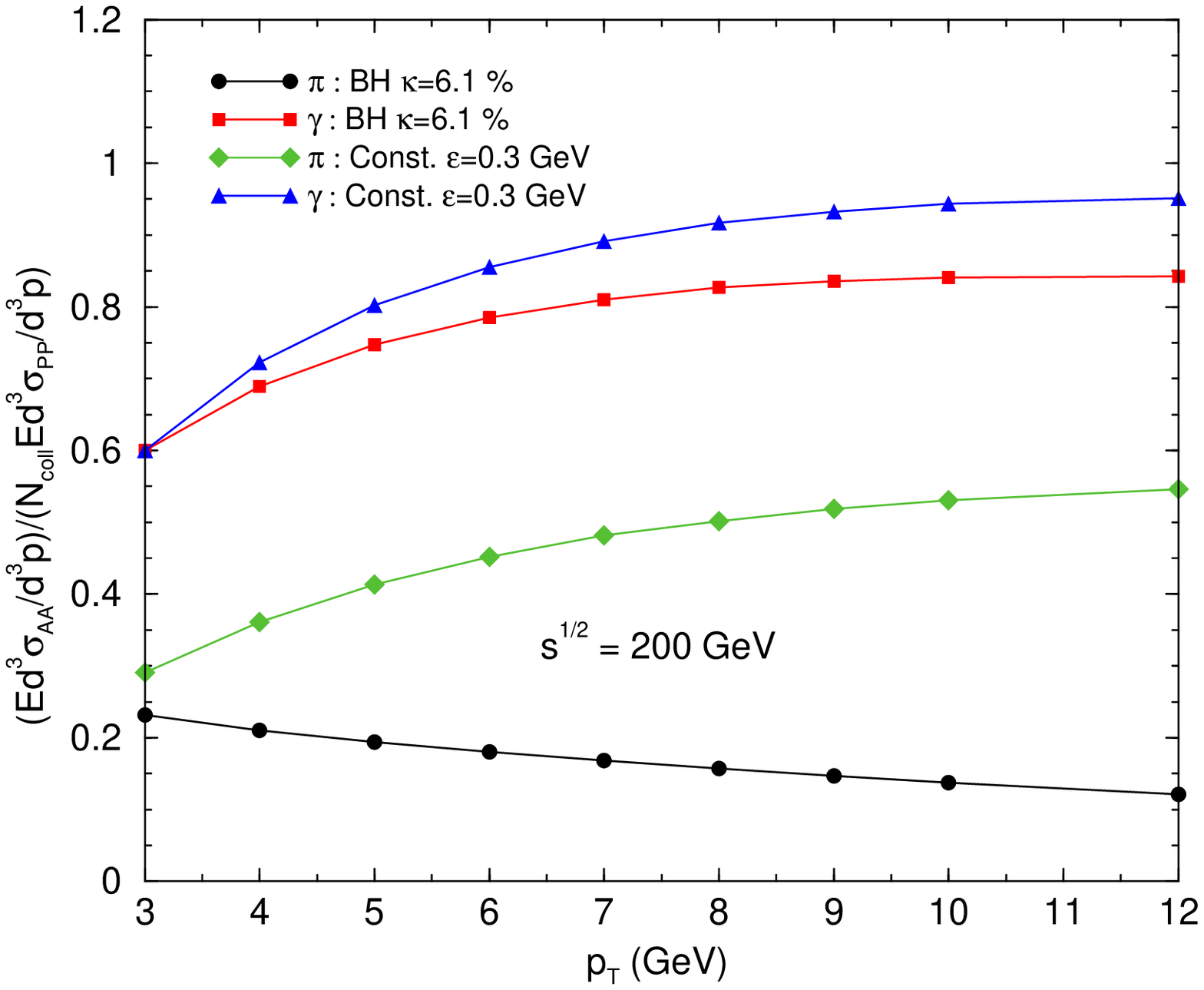}
~\vskip -9mm
\caption{Suppression of $\pi^0$ and prompt photon production in Au-Au collisions 
relative to the binary collision scaled proton-proton case, at 
$\sqrt s=200$ GeV.}
\label{fig1}
\end{minipage}
\hspace{\fill}
\begin{minipage}[t]{75mm}
\includegraphics[width=75mm]{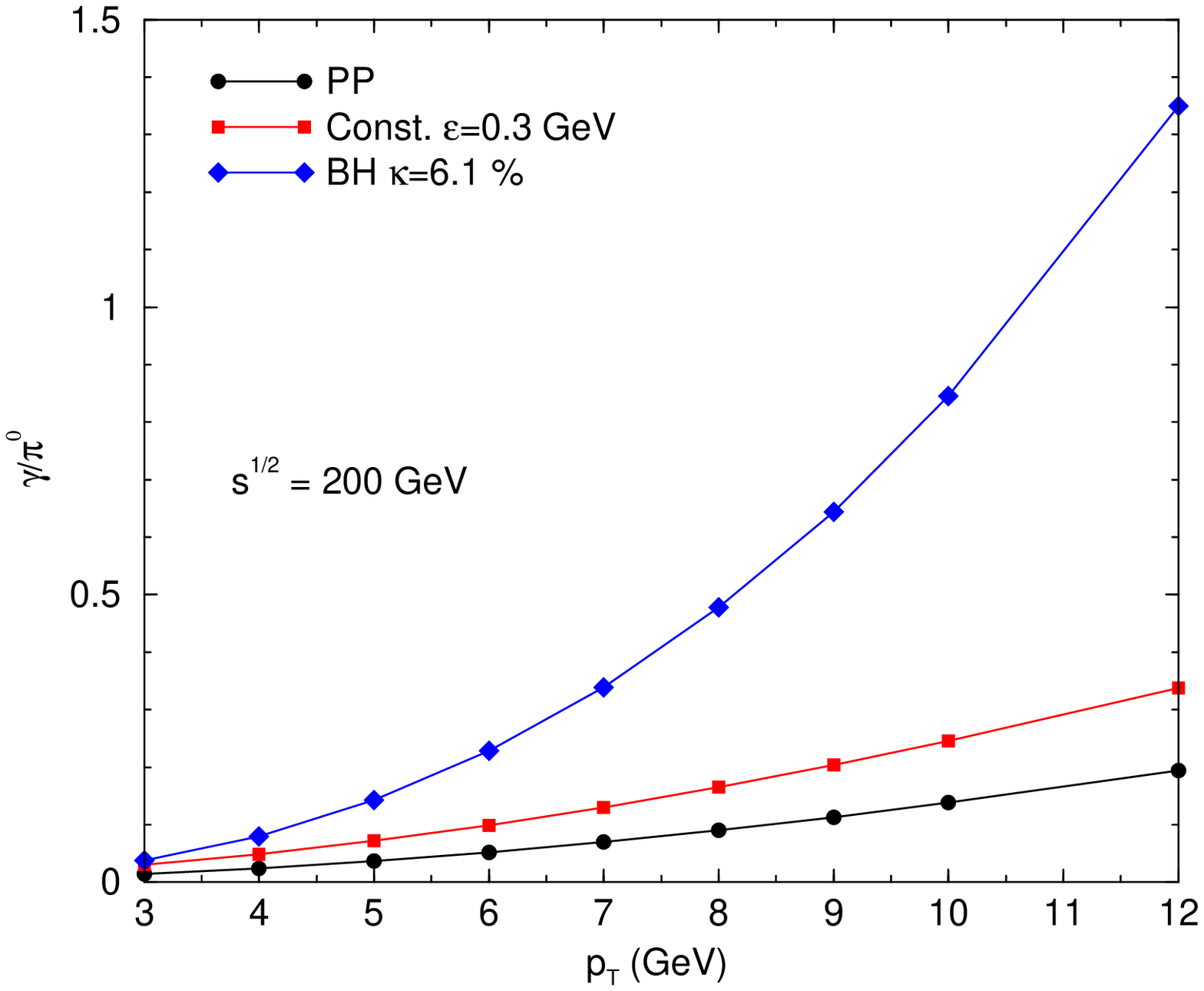}
~\vskip -9mm
\caption{The $\gamma/\pi^0$ ratio as a function of $p_T$ at 
$\sqrt s=200$ GeV.}
\label{fig2}
\end{minipage}
~\vskip -9mm
\end{figure}
 In
Fig. (3), we present $R_{AA}(p_T)$ for inclusive 
$\pi^0$ production and for prompt photon production at the LHC.  
We find that with constant parton energy loss per collision, 
 $\epsilon=1$GeV, 
pion suppression decreases from $80\%$ at $p_T=5$GeV to $20\%$ at 
$p_T=40$GeV, while for $\epsilon=0.06E$ the suppression increases 
from $70\%$ at $p_T=5$GeV to $80\%$ at $p_T=40$GeV.  
Prompt photon production in Pb-Pb collisions at the LHC is 
slightly less suppressed than $\pi^0$ production, 
for the constant energy loss, while for 
 $\epsilon=0.06E$ the suppression decreases slower, from 
$60\%$ at low $p_T$ to $30\%$ at $p_T=40$GeV.  
Nuclear shadowing effects are very small, less 
than $10\%$.  
  At this energy we note that suppression of
prompt photons is similar to the $\pi^0$ case, because at LHC
energy prompt photon production has $60\%$ contribution from 
bremsstrahlung processes, 
which are modified due to the energy loss in a similar way to the
$\pi^0$ case. We find that $\pi^0$ production is very sensitive to the 
parton energy
loss parameters. 
 
In Fig. 4 we show the ratio of prompt photons and pions at the LHC.  
We find that for constant energy loss this ratio increases slowly with 
$p_T$, similar to p-p case, while for 
 $\epsilon=0.06E$ the ratio increases rapidly approaching $0.2$ at 
$p_T=35$GeV.  

\begin{figure}[htb]
\begin{minipage}[t]{75mm}
\includegraphics[angle=0,width=75mm]{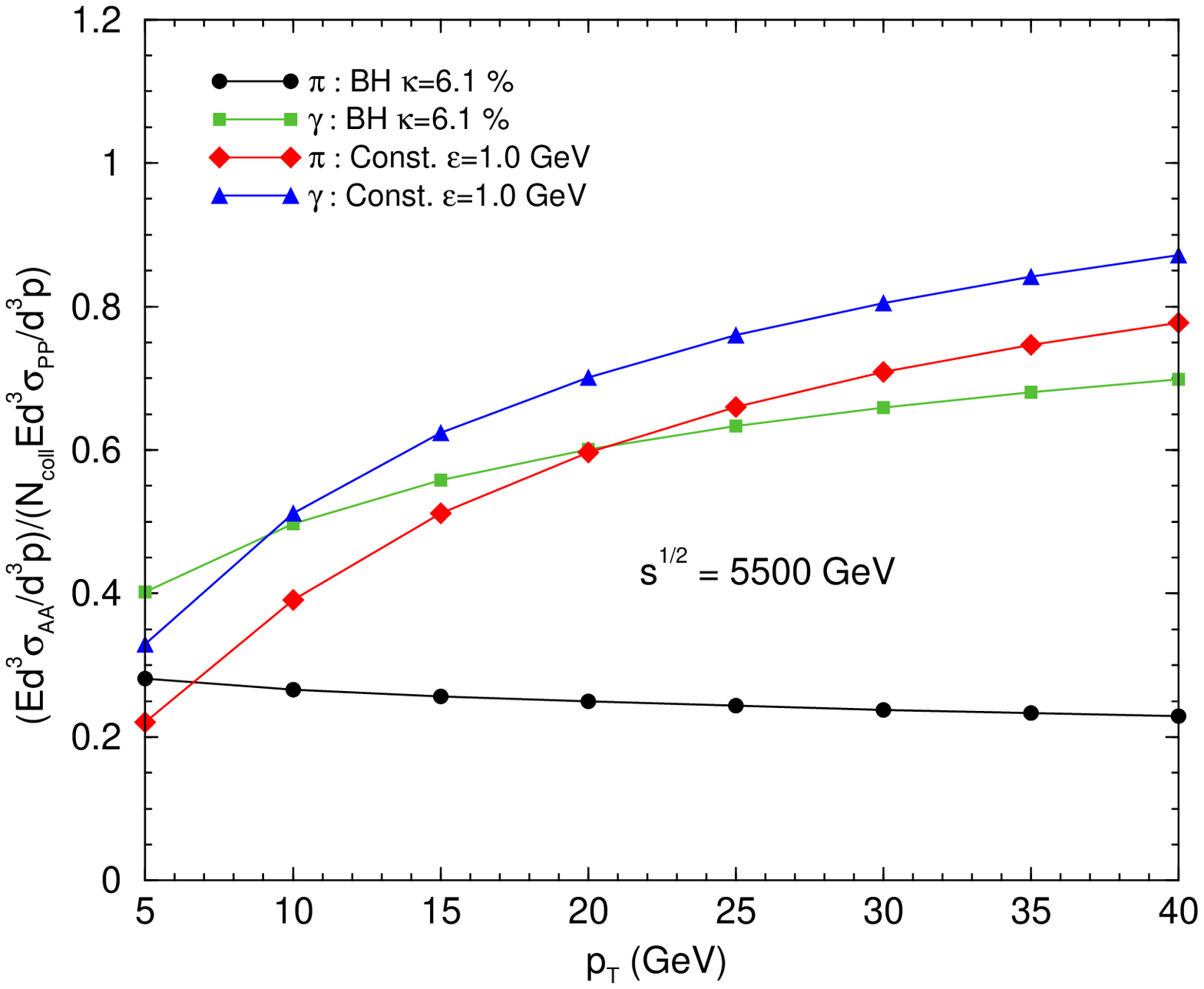}
~\vskip -9mm
\caption{Suppression of $\pi^0$ and prompt photon production in Au-Au collisions 
relative to the binary collision scaled proton-proton case, at the LHC.}
\label{fig3}
\end{minipage}
\hspace{\fill}
\begin{minipage}[t]{77mm}
\includegraphics[angle=0,width=75mm]{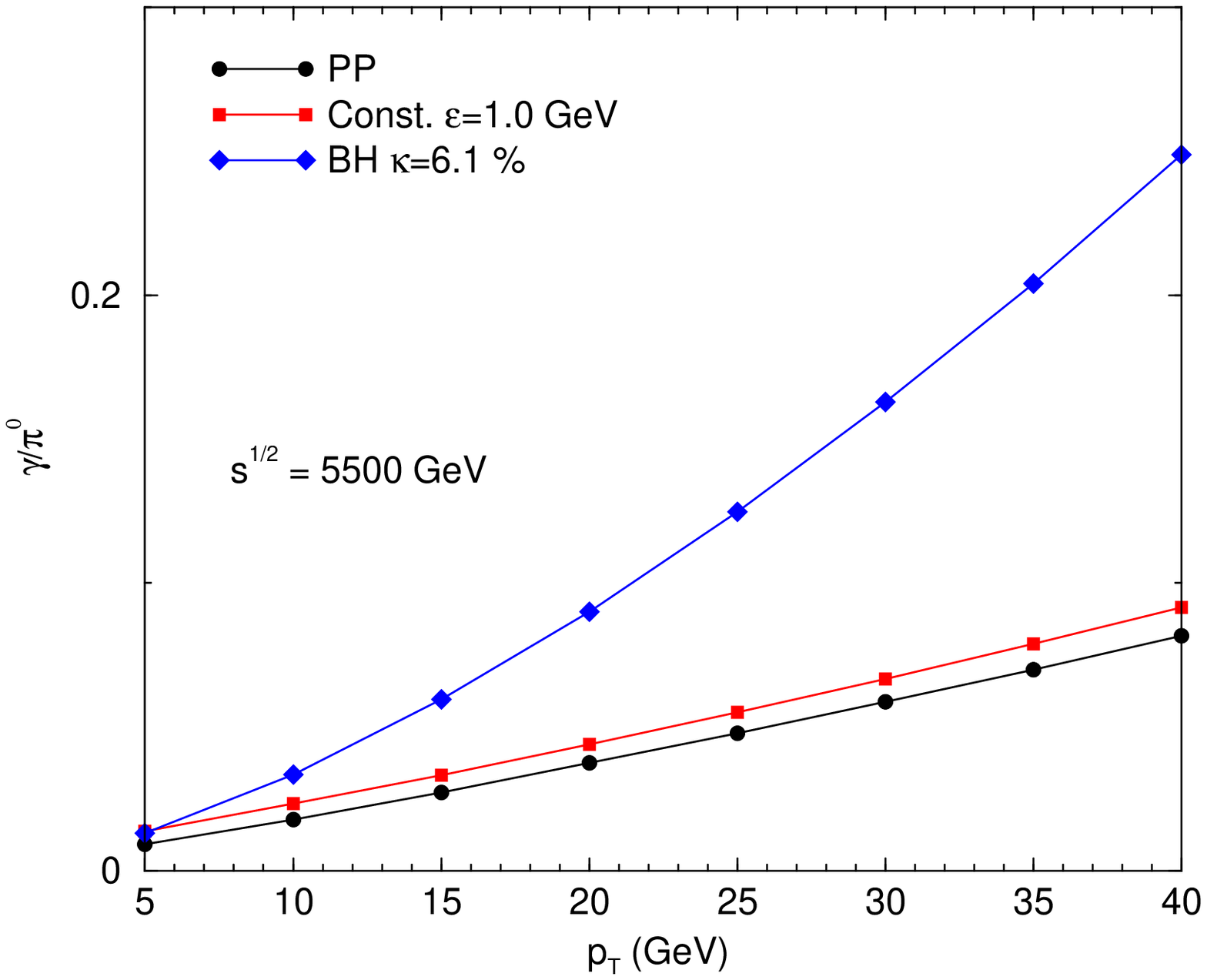}
\label{fig4}
\caption{The $\gamma/\pi^0$ ratio as a function of $p_T$ at the LHC.} 
\end{minipage}
~\vskip -9mm
\end{figure}

\section{SUMMARY}
We have calculated inclusive pion and prompt photon production
cross sections in proton-proton and in heavy-ion collisions at RHIC
and LHC energies. We have incorrporated next-to-leading order
contributions, 
initial state parton distribution functions in a nucleus and
 medium induced parton energy loss by modifying the
final state pion and photon fragmentation functions.

We find 
 the
nuclear K-factor, which signifies higher order corrections,
 to be large and $p_T$
dependent and
 the shape of the $p_T$ distribution insensitive to the
choice of scales \cite{jjs2}.
The nuclear shadowing effects are small at RHIC and LHC energies.  
 We show that $\pi^0$ suppression observed at RHIC can be attributed to 
the parton energy loss.  
We also present results for the suppression of 
 prompt photon production at RHIC and LHC, and for the ratio of 
prompt photon and $\pi^0$ cross sections, 
 of relevance to separating
 different sources of photon production. 

We are indebted to P. Aurenche and J. P. Guillet for providing us with
the fortran routines for calculating 
$\pi^0$ and photon
production in hadronic collisions and for many useful discussions.
We thank D. d'Enterria and M. Tannenbaum for many helpful
discussions and suggestions. 
 I.S. is supported in
part through U.S. Department of Energy Grants Nos. DE-FG03-93ER40792 and
DE-FG02-95ER40906. S.J. is supported in part by the Natural Sciences and
 Engineering Research Council of Canada.  
J.J-M. is supported in part by a PDF from BSA and by
U.S. Department of Energy under Contract No. DE-AC02-98CH10886.

\end{document}